\documentclass{iaus}
\usepackage{graphics}

\title[Pistol Star] 
{High-angular resolution observations of the Pistol Star}

\author[C.\ Martayan, R.\ Blomme, J.-B. Le Bouquin et al.]   
{C.\ Martayan$^{1,2}$, 
R.\ Blomme$^3$,
J.-B.\ Le Bouquin$^4$,
A.\ Merand$1$,
\\
G.\ Montagnier$^1$,
F.\ Selman$^1$,
J.\ Girard$^1$,
A.\ Fox$^1$,
D.\ Baade$^5$,
\\
Y.\ Fremat$^3$,
A.\ Lobel$^3$,
F.\ Martins$^6$,
F.\ Patru$^1$,
Th.\ Rivinius$^1$,
\\
H.\ Sana$^{7}$,
S.\ {\v S}tefl$^1$,
J.\ Zorec$^8$
\and T.\ Semaan$^2$}

\affiliation{$^1$ESO Chile; 
$^2$GEPI-Observatoire de Meudon, France; 
$^3$Royal Observatory of Belgium, Brussels, Belgium; 
$^4$Laboratoire d'Astrophysique de Grenoble, France;
\\
$^5$ESO Germany; 
$^6$GRAAL, Montpellier, France;
$^7$Amsterdam University, The Netherlands;
$^8$Institut d'Astrophysique de Paris, France}

\pubyear{2011}
\volume{272}  
\pagerange{1--2}
\jname{Active OB stars: structure, evolution, mass loss and critical limits}
\editors{C.\ Neiner, G.\ Wade, G.\ Meynet \& G.\ Peters, eds.}
\begin{document}
\maketitle

\begin{abstract}
First results of near-IR adaptive optics (AO)-assisted imaging,
interferometry, and spectroscopy of this Luminous Blue Variable (LBV) are presented.
They suggest that the Pistol Star is at least double.  If the
association is physical, it would reinforce questions concerning the
importance of multiplicity for the formation and evolution of extremely massive
stars.\footnote{based on ESO runs 085.D-0182(A), 085.D-0625(A, C)}
\keywords{stars: binaries, stars: circumstellar matter, stars: early-type, 
stars: mass loss}
\end{abstract}

\firstsection 
\section{Introducing the Pistol Star}
At the time of its formation $\sim$2 million years ago, the Pistol
Star may have been one of the most massive stars in the Milky Way
(\cite[Figer et al.\ 1998]{Figer}).  Like most other LBV's, it is
surrounded by nebulosities, which, therefore, may actually have been
ejected by the Pistol Star (cf., e.g., $\eta$ Car).  The shape of the
main gas cloud has earned the star its nick name.  Owing to its own
ejecta but mostly due to its far-away location ($\sim8$\,kpc) in the
region of the Galactic Center, observations of the Pistol Star suffer
from very high extinction (20-30 magnitudes in the optical).

Among massive stars, the fraction of binaries may approach 100\% at
the time of formation (cf.\ \cite[Mason et al. 2009]{Mason}, \cite[Sana et al. 2009]{Sana} ). In fact,
binarity may be one means to enable the formation of massive stars,
which with single-star-plus-disk symmetry encounters challenges
already at the 10-$M_{\odot}$ level while observations of LBVs imply
original masses that are larger by more than an order of magnitude.
One of the most prominent massive stars reported to be a binary is
$\eta$ Car (\cite[Damineli, Conti, \& Lopez 1997]{Damineli}).

\section{Observations}
In order to more closely examine the nature of the Pistol Star,
near-IR observations were undertaken with the VLT AO camera {\it NACO}
and the VLTI spectrograph {\it AMBER}.  The angular resolutions of
{\it NACO}, ranging from natural seeing to near the defraction limit
at $\sim$15 mas, and {\it AMBER}, 20-3 mas, are perfectly
complementary to one other. At a distance of 8\,kpc, this permits
spatial scales to be studied down to $\sim$25 AU.  The new VLT
spectrograph {\it X-Shooter} provided spectra with $R\sim5000-11000$ over
the range 0.3-2.5\,$\mu$m.  They show a vast number of nebular lines;
the stellar continuum begins to emerge at $\sim$\,800\,nm.  At the
time of the Symposium, some additional observations were pending, also
of other massive stars.

\section{First preliminary results}
\begin{figure}
\centering
\resizebox{8cm}{!}{\includegraphics{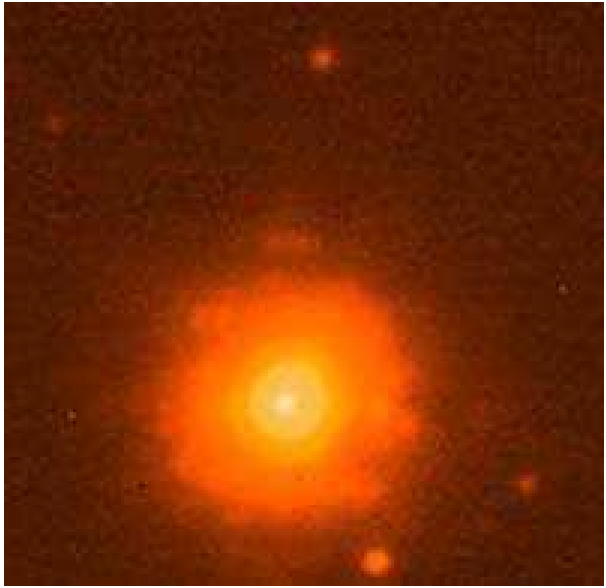}}
\caption[]{
\centering
{\it NACO} image (2.3x2.2\,arcsec) of the Pistol Star. Note the
appearance of several nearby point sources, while the Pistol Star
itself is surrounded by extended nebulosity, measuring roughly 700\,AU
across.  The structure in the nebulosity is not due to discontinuities
in the color coding but believed to be real, possibly evidencing
multiple shells.}
\label{NACO}
\end{figure}

The NACO images resolve most of the stars within 30 arcsec of the
Pistol Star into several point sources.  Fig.\ \ref{NACO} depicts the
immediate vicinity of the Pistol Star itself as observed with {\it NACO} in K band.
Apart from several point sources, it reveals an extended nebulosity.
At a distance of 8\,kpc, its diameter would correspond to more than 6000\,AU. 
\cite[Figer et al.\ 1998]{Figer} find gas velocities with a range of 
$\sim$50-100\,km/s but over a larger area.  Combining these numbers
leads to a crude estimate of the expansion age of 290 years.  If the
inference from Fig.\ \ref{NACO} of multiple shells is correct, they
would differ in age by $\sim$30\,years.

A quick preliminary inspection of the {\it AMBER} interferometry
suggests the presence of a further point source at a separation of
about 10\,mas or $\sim$\,80 AU.  For comparison: If the 5.5-year period in
$\eta$ Car is orbital, its companion would be at one-fifth or less of
this distance.  In the {\it NACO} images, the inner nebulosity looks
less structured than in the case of $\eta$ Car.  An effort will be
made to continue the observations in order to establish whether the
Pistol Star has a physical companion with effects on formation and
evolution of the system. 

\end{document}